\documentclass[conference]{IEEEtran}
\IEEEoverridecommandlockouts
\usepackage{cite}
\usepackage{amsmath,amssymb,amsfonts}
\usepackage{algorithmic}
\usepackage{graphicx}
\usepackage{textcomp}
\usepackage{xcolor}
\usepackage{multirow}
\usepackage{booktabs}
\usepackage{comment}
\usepackage{hyperref}
\makeatletter
\newcommand{\linebreakand}{%
  \end{@IEEEauthorhalign}
  \hfill\mbox{}\par
  \mbox{}\hfill\begin{@IEEEauthorhalign}
}
\makeatother

\usepackage{amsmath} 
\newcommand{\tcr}[1]{\textcolor{black}{#1}}
\def\BibTeX{{\rm B\kern-.05em{\sc i\kern-.025em b}\kern-.08em
    T\kern-.1667em\lower.7ex\hbox{E}\kern-.125emX}}
\begin{document}

\title{A Fair and Transparent Framework for Speech-Based Depression Detection: Balancing Interpretability and Performance\\
}



\author{\IEEEauthorblockN{1\textsuperscript{st} Mariel Estevez}
\IEEEauthorblockA{\textit{ViVoLab, I3A} \\
\textit{University of Zaragoza}\\
Zaragoza, Spain \\
mestevez@unizar.es}
\and
\IEEEauthorblockN{2\textsuperscript{nd} Alfonso Ortega}
\IEEEauthorblockA{\textit{ViVoLab, I3A} \\
\textit{University of Zaragoza}\\
Zaragoza, Spain \\
ortega@unizar.es}
\and
\IEEEauthorblockN{3\textsuperscript{rd} Antonio Miguel}
\IEEEauthorblockA{\textit{ViVoLab, I3A} \\
\textit{University of Zaragoza}\\
Zaragoza, Spain \\
amiguel@unizar.es}
\linebreakand 
\IEEEauthorblockN{3\textsuperscript{rd} Eduardo Lleida}
\IEEEauthorblockA{\textit{ViVoLab, I3A} \\
\textit{University of Zaragoza}\\
Zaragoza, Spain \\
lleida@unizar.es}
}

\maketitle

\begin{abstract}
While speech provides rich, non-invasive biomarkers for mental-health assessment, clinical adoption is limited by opaque models and potential demographic bias. In this work we propose a methodological framework to evaluate robustness and interpretability for automated depression detection on the extended DAIC-WOZ dataset using low-complexity machine learning baselines (RF, SVM, and MLP) chosen to mitigate overfitting and enhance generalization in combination with human-understandable acoustic features (MFCCs, eGeMAPS). To balance accuracy with clinical trust, we leverage explainability methods (LIME and SHAP) for feature selection, validating our findings with statistical significance tests and demographic fairness analyses to mitigate spurious, artifact-driven correlations. Empirical results demonstrate that an optimized subset of explainable AI (XAI)-selected features combined with an MLP architecture achieves a state-of-the-art test accuracy of 82\%. Ultimately, this work provides a transparent framework for robust and ethical assistive technologies that can be applied to any other binary task.
\end{abstract}

\begin{IEEEkeywords}
depression detection from speech, DAIC-WOZ, interpretability
\end{IEEEkeywords}

\section{Introduction}
Depression is a major public health concern, often underdiagnosed and undertreated due to structural barriers such as limited access, cost, and infrequent follow-ups, as well as the intrinsic subjectivity of clinical interviews and self-report questionnaires. This has motivated growing interest in digital biomarkers and mHealth-style screening tools that can provide objective, low-cost, and scalable support for clinicians beyond the clinic \cite{Arioz24-SRO}.
Speech is particularly attractive in this context because it is ubiquitous, non-invasive, and carries information about prosody, voice quality, and articulation, all of which may vary with mental state. Handcrafted acoustic descriptors such as Mel-Frequency Cepstral Coefficients (MFCCs) and the extended Geneva Minimalistic Acoustic Parameter Set (eGeMAPS, \cite{Eyben16-TGM}) offer a reliable and interpretable starting point, and have shown promising results. However, although these feature sets contain potentially relevant depression cues, many features may also introduce noise. Identifying the most informative subset remains challenging, since the acoustic markers most relevant for depression detection are still not fully established.
In \cite{JanardhanKumaresh2022DepressionFisherDES}, feature selection is explicitly incorporated into DAIC-WOZ \cite{Gratch14-TDA} depression detection. Starting from eGeMAPS acoustic descriptors, the authors apply redundancy reduction and Fisher score-based ranking to select a subset of features. They report 61\% accuracy using the full feature set in their baseline configuration, and up to 82\% accuracy after applying their best feature-selection strategy, highlighting the importance of selecting robust and informative acoustic descriptors. \tcr{Nevertheless, as noted in \cite{Zhangetal2024}, although manual feature extraction has demonstrated efficacy in depression detection, selecting appropriate tools often requires substantial domain expertise.}

More complex and opaque representations can be extracted from self-supervised learning (SSL) speech encoders originally developed for tasks such as automatic speech recognition and speaker verification. Although these systems achieve strong performance in several speech-processing tasks, they often behave as black boxes and provide limited insight into the basis of their decisions, which is problematic in high-stakes clinical settings where transparency and trust are essential \cite{Ibrahimov24-EAI}. Recent evaluations of clinical speech tools caution against the blind use of large-scale feature extraction and black-box embeddings \cite{Xu22-COA, Choi23-CEO}. Such approaches may achieve high benchmark performance while failing to provide the clinical interpretability required for laryngeal or mental health diagnostics.
Moreover, although SSL-based deep learning systems trained on clinical-interview corpora such as DAIC-WOZ are widely used for depression detection, their reliability, reproducibility, and methodological rigor remain uncertain, as discussed in \cite{app16010422, Yehetal2026}. Recent evidence shows that models can exploit biases correlated with interviewer behavior, such as targeted prompts localized in specific interview regions, thereby inflating apparent performance without learning generalizable patient cues \cite{Burdisso24-DWO}. Similarly, although several SSL-based systems report accuracies above 90\% \cite{Xu2025DepressionDM}, \cite{Yehetal2026} argues that speaker leakage may drive these inflated metrics. In no-leakage settings, reported results are only slightly better than those of simpler models \cite{Yehetal2026, Othmanietal2021, Dumpalaetal2021, Tasnim2019, JanardhanKumaresh2022DepressionFisherDES, Yalamanchili2020icETITE_Depression}, while retaining the additional complexity associated with deep SSL-based pipelines.

Fairness is another key concern. Evidence from related biomedical speech-classification settings shows that strong pooled performance may coexist with severe disparities across age and gender groups, motivating group-wise analysis and cost-sensitive evaluation rather than relying only on aggregate metrics \cite{Estevez24-BGM}.
Taken together, these observations motivate depression detection pipelines that (i) prioritize human-understandable features, (ii) explicitly test robustness against spurious shortcuts, and (iii) quantify demographic sensitivity under clinically meaningful operating points.

In this work, we take a methodological step toward an explainable Mental Illness Assessment System based only on voice by studying the robustness and interpretability of acoustic features for automated depression detection on the extended DAIC-WOZ dataset \cite{Ringeval19-AVEC}. We focus on low-complexity models, including Random Forests (RF), Support Vector Machines (SVM), and Multilayer Perceptrons (MLP), to isolate the predictive content of human-interpretable descriptors such as $f_0$-related measures, perturbation cues, and spectral/cepstral features.
We compare several feature-selection strategies, including LIME \cite{Ribeiro16-WSI}, SHAP \cite{Lundberg17-AUN}, and statistical selection. \tcr{These methods are inherently interpretable and, consequently, capable of identifying significant features. Furthermore, owing to the adopted methodological framework, they can be readily applied without requiring extensive expert knowledge.} We further assess statistical significance through permutation tests and interviewer-participant consistency checks, in order to evaluate whether 
the use of the interviewer's voice can artificially enhance the performance of the model. The proposed procedure is independent of the unique characteristics of this specific dataset or task; hence, it can be applied to any binary-labeled system that uses a list of features and RF, SVM or MLP baselines. We also perform demographic fairness analyses using the Normalized Expected Cost (NEC), emphasizing clinically aligned error trade-offs. Finally, we show that a simple and clinically manageable model, combined with an appropriate subset of acoustic features, can achieve state-of-the-art results and outperform more complex alternatives.

\section{Methods}
In this section, we outline the experimental setup used to train the classification systems and evaluate their performance. Specifically, we detail the datasets, the feature extraction and selection processes, the classification models, the statistical tests, and the evaluation metrics employed.

\subsection{Dataset and Preprocessing}
For this study, we utilized the audio recordings and transcriptions from the DAIC-WOZ dataset, along with its metadata. Binary depression labels were derived using the PHQ-8 scores, where a score equal to or larger than 10 indicates that the patient is suspected of having Major Depression Disorder (MDD) \cite{Sadeghi2024HarnessingMultimodalDepression}. We also included the new participants from the E-DAIC corpus dataset, for which we automatically isolated participant speech using WhisperX \cite{Bain23-WTA}, leveraging Whisper \cite{Radford23} for transcription and Pyannote \cite{Bredin23} for accurate speaker diarization.

To ensure data quality, we applied automated heuristic filtering to correct diarization errors. We excluded extraneous initial dialog occurring before the agent's standardized greeting ("Hi, I'm Ellie") and reassigned utterances to the interviewer if the ASR transcription contained question or exclamation marks but lacked first-person pronouns (e.g.,"I" or "me"). The complete dataset was randomly divided into an 80\% training set and a 20\% test set, assuring each participant is only present in one set to avoid speaker id leakage. While the E-DAIC corpus provides standard data partitions, these static splits limit the ability to measure feature stability across varying data distributions. Because our primary objective is to evaluate the robustness and interpretability of acoustic features rather than merely maximizing benchmark accuracy, we purposely merged the corpus and performed a custom random partition to mitigate severe domain shifts between the official splits; this design choice establishes an acoustic balance across sets, preventing models from exploiting partition-specific environmental artifacts and ensuring that our interpretability metrics reflect genuine clinical biomarkers. The test set was strictly held out from all training and feature selection processes, and was used solely for computing the final metrics and conducting statistical tests. The ids list per set can be found in \url{https://anonymous.4open.science/r/DAIC-WOZ_interpretability_framework-2D26}.

\subsection{Audio Preparation and Feature Extraction}
All audio files were converted to mono and resampled to a 16 kHz sampling rate. We filtered out sessions 451, 458, 480 (where Ellie is absent), and 620 due to poor diarization quality.

The audio recordings were segmented into turn-taking intervals; a turn begins when a speaker starts speaking and ends when the interlocutor takes the floor. For each participant turn, we calculated the response latency (the delay before responding to the previous speaker) and the syllable count from the WhisperX  transcripts using spaCy \cite{Honnibal20-SPI}. We applied the Silero Voice Activity Detection (VAD) model \cite{Silero24-VAD} to mask non-speech intervals. This allowed us to extract the following temporal features: speech duration (duration of detected speech), silence duration, speech rate (syllables per VAD-active speech time), articulation rate (syllables per speech time), and the ratio of speech time to total turn duration. Additionally, we computed the response latency following each question (\texttt{response\_dur}).

Acoustic processing was performed using TorchAudio \cite{Yang21-TAB} to extract custom Mel-Frequency Cepstral Coefficients (MFCCs). We computed frame-level MFCCs using \texttt{mfcc\_transform}. To prevent artificial discontinuities in the temporal dynamics, we first computed the first- and second-order derivatives ($\Delta$ and $\Delta\Delta$) of  the MFCCs and then applied the VAD mask to remove non-speech frames. Alongside the MFCCs, we extracted additional acoustic low-level descriptors (LLDs), including zero-crossing rate, spectral flux, and extreme ranges. For all these descriptors, the mean and standard deviation were computed. Finally, we used openSMILE (open-source Speech and Music Interpretation by Large-space Extraction, \cite{Eyben10-OTM}) to extract the 88 eGeMAPS feature set for each turn without using the VAD mask since silence time is often correlated with signs of depression in eGeMAPS features.

Operating under the assumption that depression-related markers are more reliably captured over sustained speech, we retained only the turns longer than 4 seconds and discarded those containing more than five null feature values. Feature aggregation was performed at the speaker level by averaging the turn level statistics, resulting in a single feature vector per speaker.

\subsection{Feature Selection Strategies}
Our general procedure involves training various models using distinct feature subsets to ascertain their real, non-spurious importance. Feature selection was strictly performed on the training set using Stratified K-Fold cross-validation with a fixed seed. To derive these subsets, we employed five distinct selection strategies, each identifying the 15 most relevant features:

\begin{enumerate}
    \item \textbf{Baseline}: To simulate an uninformative feature space and verify if high dimensionality introduces redundant noise rather than predictive value, a subset of 15 randomly selected features was utilized as a baseline control.
    \item \textbf{Statistical Selection (sigst):} We compared the feature distributions between classes within the training set, following \cite{Sheskin20-HAP}. Normality was assessed using the Shapiro-Wilk test, and homoscedasticity via Levene's test. Depending on these assumptions, we applied the Student's t-test (for normal, equal variances), Welch's t-test (for normal, unequal variances), or the Mann-Whitney U test (for non-normal distributions). Features were ranked by p-value in ascending order, and the top 15 were selected for exploratory purposes, acknowledging the caveat that p-values from different tests are not strictly comparable.
    \item \textbf{SYSstem Importance Selection (SYSsel):} We estimated feature importance using permutation importance \cite{breiman2001} (via scikit-learn \cite{pedregosa11}) during training within the K-fold cross-validation setup and based on accuracy. We aggregated the importances by averaging across folds, selecting the top 15 features.
    \item \textbf{LIME:} The top 15 features were selected based on their global importance calculated using the Local Interpretable Model-agnostic Explanations (LIME) algorithm in an Out-of-Fold (OOF) K-fold cross-validation. To derive global feature relevance, we implemented a systematic aggregation of local surrogate coefficients by calculating the mean absolute importance across all out-of-fold training samples.
    \item \textbf{SHAP:} Similar to LIME, we selected the 15 most important features based on the SHapley Additive exPlanations (SHAP) algorithm using the OOF training set.

\end{enumerate}

\subsection{Classification and Calibration Models}
We opted for basic binary classifiers (RF, SVM and MLP) implemented via the scikit-learn library to ensure simplicity and reproducibility. For the SVM and MLP models, the input features were standardized to have a zero mean and unit variance. 

The specific configurations for each classifier were defined as follows:
\begin{itemize}
    \item \textbf{RF:} Configured with 10 trees (estimators), an unbounded maximum depth, and a minimum of 2 samples required to be at a leaf node. Class weights were balanced to automatically adjust for the class frequencies in the training data.
    \item \textbf{SVM:} Implemented using a Radial Basis Function (RBF) kernel with a regularization parameter ($C$) set to 10 and a scale-adjusted kernel coefficient ($\gamma$). Similar to the RF, class weights were set to balanced.
    \item \textbf{MLP:} Designed with two hidden layers containing 1024 neurons each, utilizing the Rectified Linear Unit (ReLU) activation function. The network was optimized using the Adam solver with an L2 penalty ($\alpha$) of $10^{-4}$, an adaptive learning rate, and trained for a maximum of 500 iterations. We chose parameters to mimic \cite{Gheorghe19-UDN}.
\end{itemize}

To ensure robust statistical evaluation, we employed a Stratified K-Fold cross-validation (5 folds) across 10 different random seeds, which controlled both the system initialization parameters and the cross-validation splits. For each system, seed and fold, the predictions of the test set were calibrated using Platt Scaling. For SVM we calibrated margins as in \cite{Platt99-POF}. The calibrated probabilities across folds were then fused using a logit mean \cite{Genest86-CPD}. 
Unlike the standard arithmetic mean, fusing scores in the logit (log-odds) space 
allows for a linear combination of the models' evidence. 

\subsection{Performance Metrics}
In this work, we evaluate system performance using the Expected Cost (EC) and its normalised version, the Normalised Expected Cost (NEC). For a binary task, taking the costs for making a correct decision as zero, the EC reduces to
$\mathrm{EC} = c_{FN}P_{D}R_{FN} + c_{FP}P_{H}R_{FP}$ ,
where $c_{FN}$ and $c_{FP}$ are the costs of false negatives and false positives respectively, $R_{FP}$ is the false positive rate, $R_{FN}$ is the false negative rate, $P_{D}$ is the
proportion of samples belonging to a patient with MDD and $P_H$
is the proportion of samples that belongs to a control one. 

These costs parameters $c$ are highly application-dependent and should be set by clinical domain experts to maximize the real-world utility and safety of the screening tool. For our fairness analysis across genders, we set equal costs for false positives and false negatives ($c_{FN} = c_{FP} = 1$). Under this configuration, the EC effectively reduces to the total error rate ($1 - \mathrm{ACC}$). We normalize this value by the expected cost of a naive system: one that has no access to the input samples and always predicts the most frequent class. The resulting NEC offers an intuitive interpretation: a NEC value of 1 represents a performance equivalent to the random/naive baseline, and values below 1 indicate the degree of improvement the model achieves over this baseline.

\tcr{saco todo el parrafo este? It is pertinent to note that, for reasons of direct comparability with ACC and interpretative simplicity, this study used a NEC with equal costs. However, given the clinical criticality inherent in depression detection, it is highly recommended for future deployments to use weighting schemes that penalize false negatives more asymmetrically. Likewise, the final selection of a clinical model must contemplate an ethical balance between the global optimization of the NEC and its variance across different demographic strata.}

\subsection{Statistical Tests}
We report three complementary significance analyses.
\textbf{(A) Permutation test:} we compare the observed mean NEC (averaged over seeds) against a null distribution obtained from $B$ label-shuffled runs, and compute a one-sided $p$-value.
\textbf{(B) Consistency (participant vs. interviewer):} we compute NEC for participant-speech and Ellie-speech systems under the same configuration and apply a paired Wilcoxon signed-rank test across seeds to test whether the median NEC difference is zero.
\textbf{(C) Class separation:} using the calibrated per-subject score log-odds $s$, we test the \emph{non-depressed} score distribution $s \mid (y=0)$ against the \emph{depressed} score distribution $s \mid (y=1)$ with a Mann-Whitney $U$ test per seed and combine $p$-values across seeds with Fisher's method. NEC and $s$ are derived from foldwise calibration and logit-mean aggregation. We follow the procedures outlined in \cite{Sheskin20-HAP} for the tests selection.

\section{Results and Discussion}

As a result of the predictions on the isolated test set, we obtained the calibrated probabilities for each system. These estimates were used to calculate performance metrics and to extract both means and standard deviations across the ten random seeds. Table \ref{tab:resultados} presents these results, along with the $p$-values corresponding to each of the three statistical analyses applied.

\begin{table}[h!]
\footnotesize
\centering
\caption{Performance metric results by system and feature subset. Superscripts $A$, $B$, and $C$ denote statistical significance ($p < 0.05$) against the baseline for Test A (Permutation), Test B (Consistency), and Test C (Class separation), respectively. In each case, SYSsel refers to its corresponding system.}
\label{tab:resultados}
\resizebox{\columnwidth}{!}{%
\begin{tabular}{llcc}
\toprule
\textbf{System} & \textbf{Group} & \textbf{ACC (Mean $\pm$ SD)} & \textbf{NEC (Mean $\pm$ SD)} \\ \midrule
\multirow{5}{*}{RF}  & Baseline$^{B}$  & $0.66 \pm 0.01$ &                        $1.01 \pm 0.02$ \\
                     & sigst$^{C}$   & $0.66 \pm 0.01$ & $1.01 \pm 0.04$ \\ 
                     & SYSsel & $0.67 \pm 0.01$ & $0.99 \pm 0.03$ \\
                     & LIME$^{C}$    & $0.65 \pm 0.01$ & $1.04 \pm 0.03$ \\
                     & SHAP$^{C}$    & $0.66 \pm 0.01$ & $1.03 \pm 0.03$ \\
                     \midrule
\multirow{5}{*}{SVM} & Baseline$^{B}$  & $0.67 \pm 0.00$ & $1.00 \pm 0.00$ \\
                     & sigst$^{BC}$  & $0.69 \pm 0.01$ & $0.94 \pm 0.04$ \\ 
                     & SYSsel$^{B}$& $0.64 \pm 0.01$ & $1.09 \pm 0.03$ \\
                     & LIME$^{ABC}$  & $0.71 \pm 0.03$ & $0.87 \pm 0.09$ \\
                     & SHAP$^{ABC}$  & $0.69 \pm 0.01$ & $0.92 \pm 0.03$ \\
                     \midrule
\multirow{5}{*}{MLP} & Baseline$^{B}$  & $0.67 \pm 0.00$ & $1.00 \pm 0.00$ \\
                     & sigst$^{ABC}$ & $0.72 \pm 0.02$ & $0.85 \pm 0.05$ \\ 
                     & SYSsel$^{ABC}$& $0.68 \pm 0.01$ & $0.95 \pm 0.03$ \\
                     & LIME$^{ABC}$  & $\mathbf{0.82 \pm 0.02}$ & $\mathbf{0.53 \pm 0.07}$ \\
                     & SHAP$^{BC}$   & $0.67 \pm 0.02$ & $0.98 \pm 0.07$ \\
                     \bottomrule
\end{tabular}
}
\end{table}

\subsection{Performance Analysis and Class Discrimination}

As observed, accuracy (ACC) values remain remarkably similar across most configurations, staying close to the prior probability of the majority class ($0.69$), with the exception of the best model, where it achieves 82\%. The baseline accuracy values are comparable with those reported in \cite{JanardhanKumaresh2022DepressionFisherDES} for RF and in \cite{Yalamanchili2020icETITE_Depression} for RF and SVM.
In contrast, the Normalized Expected Cost (NEC) values experience more drastic variations. Notably, it decreases from $1.0$ or above (equivalent to a random/naive classifier or worst) to nearly half ($0.53$) for the best-performing model. This metric provides a robust assessment of whether the model achieves a meaningful predictive gain over a naive baseline based on class priors.

\begin{figure}[ht]
    \centering
    \includegraphics[width=0.8\columnwidth]{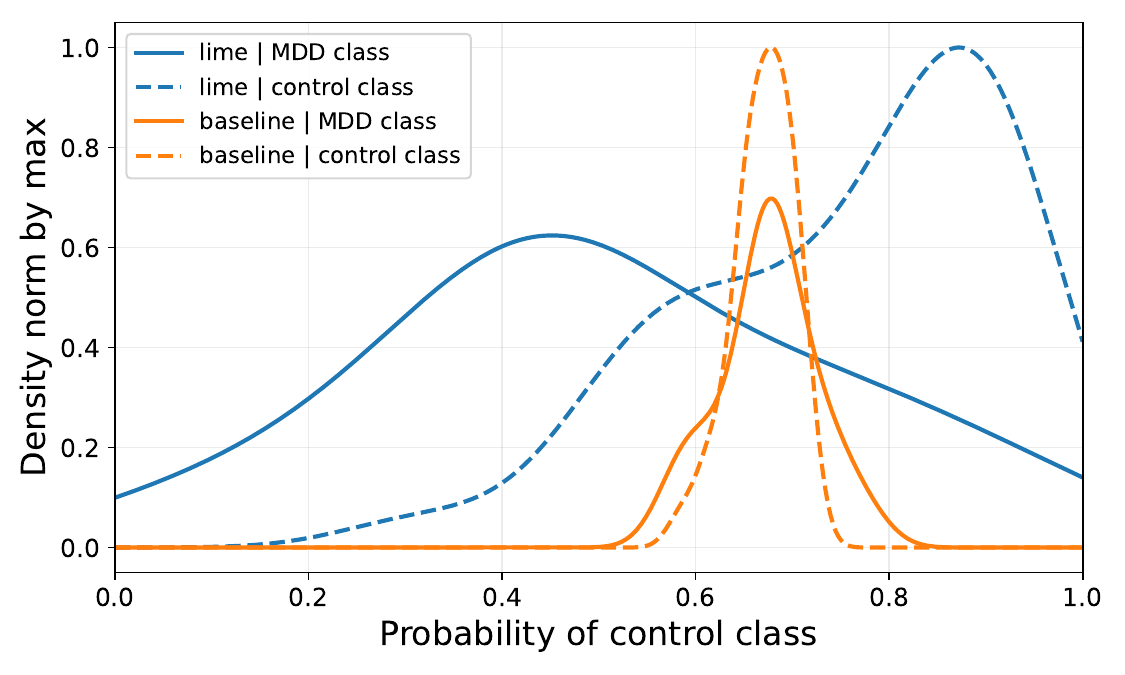}
    \caption{MLP calibrated probability distributions for both classes for the best-performing system (LIME) and the baseline. Solid lines represent the MDD class, linked to PHQ8 $\geq 10$ scores, while dashed lines indicate the control group class. It is observable that the distributions for the baseline system coincide, with their peaks centered around the prior probability.}
    \label{fig:distributions}
\end{figure}

The results indicate that the MLP architecture provides the best NEC values, particularly standing out when using the feature subsets selected through LIME and \textit{sigst}. This behavior is desirable, as these features are selected following theoretical and interpretability foundations that justify their causal relevance. This contrasts sharply with algorithmic importance methods (\textit{SYSsel}), which prioritize metric optimizations that may arise from spurious correlations. As can be seen in Figure \ref{fig:distributions} the MLP-LIME system demonstrates superior class discrimination, with clearly separated probability mass for each group.

In Test A, only a small subset of models, specifically those with the lowest NEC, achieved statistical significance ($p<0.05$). For the remaining systems, the fact that label shuffling did not significantly degrade performance suggests they failed to capture genuine clinical indicators of depression.
Test B is less restrictive than A, and all SVM and MLP systems achieved statistically different values ($p < 0.05$), indicating that they are able to identify participant characteristics related to the depression label that should not be present in Ellie's voice. 
Notice that RF-baseline also presents $p<0.05$,  due to the fact that the system performs systematically worse on the agent's voice than on the participant's.
Correspondingly, the results of Test C reveal that systems with a $p$-value $> 0.05$ are predominantly related to baseline feature subsets or algorithmic selections (e.g., SYSsel for RF and SVM). This implies that, in these systems, the distributions of the two classes are statistically indistinguishable. Since the distributions of both classes lie on the same side of the decision threshold, most instances of the majority class (control subjects) are classified correctly. Thus, the ACC is artificially around the majority class prior. However, the NEC effectively penalizes this behavior, allowing for the isolation of truly discriminative models. Exceptionally, the MLP-SHAP configuration show performance close to the naive system, which is attributed to an extensive tail in the distribution of the MDD class toward the incorrect side of the threshold, severely penalizing the expected cost. In conclusion, the tests evidence the superiority of the MLP-LIME and MLP-sigst models, which consistently yielded the lowest NEC values and $p$-values $< 0.05$, demonstrating their ability to capture the underlying information of the patient's voice in a robust and non-stochastic manner.

\subsection{Analysis of the features in the MLP-LIME selection}
\begin{figure}[ht]
    \centering
    \includegraphics[width=0.8\columnwidth]{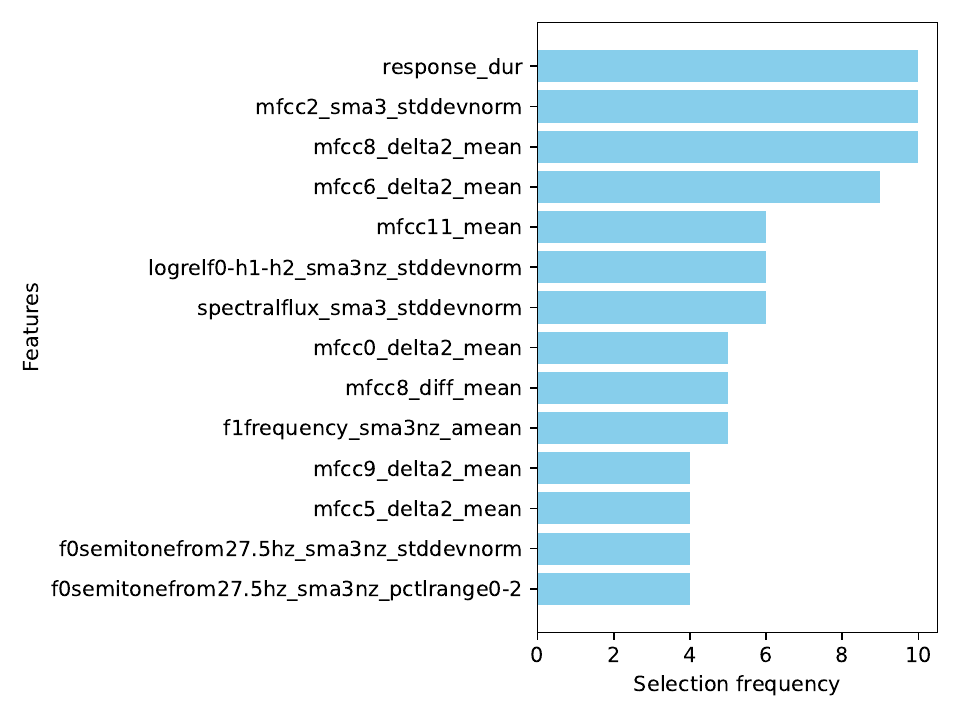}
    \caption{Histogram of feature occurrence counts across the 12 distinct machine learning systems, showing the most frequently selected audio features. \texttt{sma3} and \texttt{sma3nz} subscripts refer to eGeMAPS features.}
    \label{fig:histograma}
\end{figure}

In this section we do a review on the interpretability of the features selected from the models. Figure \ref{fig:histograma} shows the most frequent features appearing in the 12 models, combining RF, SVM and MLP systems with sigst, SYSsel, LIME and SHAP selection strategies. The selected features capture complementary speech dimensions potentially affected in depression, including response timing, cepstral/spectral-envelope dynamics, pitch variability, voice quality, spectral flux, and formant-related articulation. Together, they suggest that depression-related information is distributed across temporal, prosodic, spectral, phonatory, and articulatory patterns rather than concentrated in a single acoustic marker. The most relevant features are discussed in more detail below.

Since the MLP-LIME selection strategy achieved the highest performance in this work, we focus the following interpretability discussion on its features, analyzing them based on their importance in the LIME selection set and their training set distributions computed for the statistical selection. Table \ref{tab:LIMEselected_features} presents the list of these features, including their order of importance from the LIME algorithm and descriptor type. In this scheme, several prosodic, spectral, and temporal features consistently emerged as highly discriminative.
\begin{table}[htbp]
\caption{Selected features from the MLP-LIME model ordered by relevance}
\begin{center}
\tiny
\begin{tabular}{|c|c|c|}
\hline
\textbf{Importance order} & \textbf{Descriptor type} & \textbf{Feature} \\
\hline
1 & psychomotor & \texttt{response\_dur} \\
\hline
2 & cepstral / spectral-envelope & \texttt{mfcc6\_delta2\_mean} \\
\hline
3 & cepstral / spectral-envelope & \texttt{mfcc2\_sma3\_stddevnorm} \\
\hline
4 & cepstral / spectral-envelope & \texttt{mfcc3\_sma3\_stddevnorm} \\
\hline
5 & cepstral / spectral-envelope & \texttt{mfcc8\_delta2\_mean} \\
\hline
6 & paralinguistic acoustic & \texttt{f1frequency\_sma3nz\_amean} \\
\hline
7 & paralinguistic acoustic & \texttt{f0semitonefrom27.5hz\_sma3nz\_pctlrange0-2} \\
\hline
8 & cepstral / spectral-envelope & \texttt{mfcc0\_delta2\_mean} \\
\hline
9 & paralinguistic acoustic & \texttt{f0semitonefrom27.5hz\_sma3nz\_meanrisingslope} \\
\hline
10 & cepstral / spectral-envelope & \texttt{mfcc9\_delta2\_mean} \\
\hline
11 & paralinguistic acoustic & \texttt{f0semitonefrom27.5hz\_sma3nz\_stddevnorm} \\
\hline
12 & cepstral / spectral-envelope & \texttt{mfcc2\_mean} \\
\hline
13 & cepstral / spectral-envelope & \texttt{mfcc10\_std} \\
\hline
14 & cepstral / spectral-envelope & \texttt{mfcc3\_delta\_mean} \\
\hline
15 & paralinguistic acoustic & \texttt{f3bandwidth\_sma3nz\_stddevnorm} \\
\hline
\end{tabular}
\label{tab:LIMEselected_features}
\end{center}
\end{table}

From a temporal perspective, \texttt{response\_dur} suggests that individuals with depression may exhibit longer response latencies, which is consistent with clinical observations of psychomotor retardation.
The selected MFCC-based descriptors point to differences in short-term spectral-envelope dynamics between groups. Features such as \texttt{mfcc6\_delta2\_mean}, \texttt{mfcc8\_delta2\_mean}, \texttt{mfcc9\_delta2\_mean}, \texttt{mfcc0\_delta2\_mean}, and \texttt{mfcc3\_delta\_mean} capture first- and second-order temporal changes in the cepstral representation, suggesting that the classifier is sensitive not only to the average spectral shape of the voice but also to how this shape evolves over time. This may reflect altered articulatory transitions, differences in vocal stability, or changes in the local dynamics of speech production. 
Variability-based MFCC descriptors, including \texttt{mfcc2\_sma3\_stddevnorm}, \texttt{mfcc2\_sma3\_stddevnorm}, \texttt{mfcc3\_sma3\_stddevnorm} and \texttt{mfcc10\_std}, may indicate altered or more heterogeneous spectral structure in depressed speech.
Pitch-related features provide additional evidence that prosodic dynamics are relevant. The presence of \texttt{f0semitonefrom27.5hz\_sma3nz\_pctlrange0-2}, \texttt{f0semitonefrom27.5hz\_sma3nz\_stddevnorm}, and \texttt{f0semitonefrom27.5hz\_sma3nz\_meanrisingslope} indicates that the model uses information related to pitch range, pitch variability, and the steepness of rising $F_0$ movements. However, the pitch range and pitch variability distributions do not appear to support a simple monotonic-speech explanation and thus, the $F_0$ features should be interpreted as evidence of altered prosodic dynamics rather than uniformly reduced pitch variability.
Finally, the inclusion of \texttt{f1frequency\_sma3nz\_amean} and \texttt{f3bandwidth\_sma3nz\_stddevnorm} indicates that formant-related information also contribute to the model. These descriptors reflect differences in articulatory configuration, vocal-tract resonance, or phonetic content. 
Overall, this feature set suggests that the classifier relies on a combination of response timing, prosodic behavior, spectral-envelope dynamics, and formant-related information, rather than on a single acoustic marker of depression.
Low et al. \cite{Low2020_LIO2_354} summarize that depression is frequently reflected in prosodic/phonatory markers (e.g., $f_0$ level/range/variability, voice-quality cues) and temporal slowing, while Cummins et al. \cite{Cummins2023_JAD_MultilingualMarkers} report the most robust cross-language associations with depressive symptoms for global motor/output measures such as speech rate, articulation rate, and intensity. 
Our features are broadly consistent with these prior evidence on depression-related speech changes. Thus, we conclude that the LIME-based feature selection set is highly interpretable in terms of depression.

\subsection{Demographic Fairness Analysis}

Finally, the fairness evaluation by age-group and gender evidence performance disparities depending on the architecture employed. In Table \ref{tab:comprehensive_demographics} we show the NEC for each group in the MLP system, giving its best metrics and statistical relevance in most systems. Although the MLP-LIME model yield the lowest NEC values globally, a clear bias is observed in favor of young (0-30 yo) and adult (30-50 yo) male participants and older female adults (45+). This model yields lower NEC values for males in general and, taking into account dispersion, comparable NEC values for the three age groups. Notice anyway that the number of participant in each group is small and unbalanced, which can lead to bias coming from poor statistics rather than real data behavior.

\begin{table}[h]
\footnotesize
\centering
\caption{Mean and standard deviation Normalized Expected Cost (NEC) across demographic combinations for the MLP model. Results are aggregated by intersectional groups (Age and Gender) as well as marginal groups (Gender only, Age only). The lowest NEC per demographic row is highlighted in bold.}
\label{tab:comprehensive_demographics}
\resizebox{\columnwidth}{!}{
\begin{tabular}{llcccc}
\hline
\textbf{Age} & \textbf{Gender} & \textbf{SYSsel} & \textbf{LIME} & \textbf{SHAP} & \textbf{sigst} \\ \hline
\multicolumn{6}{c}{\textbf{Intersectional Groups}} \\ \hline
\multirow{2}{*}{0-30}  & Fem (N=6)  & $1.03 \pm 0.19$ & $\mathbf{0.67 \pm 0.00}$ & $\mathbf{0.67 \pm 0.00}$ & $0.90 \pm 0.16$ \\
                       & Male (N=7) & $0.67 \pm 0.00$ & $\mathbf{0.33 \pm 0.00}$ & $0.50 \pm 0.18$ & $0.67 \pm 0.00$ \\ \hline
\multirow{2}{*}{30-45} & Fem (N=8)  & $1.00 \pm 0.00$ & $0.83 \pm 0.17$ & $1.48 \pm 0.08$ & $\mathbf{0.80 \pm 0.11}$ \\
                       & Male (N=6) & $0.67 \pm 0.00$ & $\mathbf{0.37 \pm 0.19}$ & $0.77 \pm 0.32$ & $0.67 \pm 0.00$ \\ \hline
\multirow{2}{*}{45+}   & Fem (N=10) & $1.00 \pm 0.00$ & $\mathbf{0.37 \pm 0.11}$ & $0.47 \pm 0.17$ & $0.80 \pm 0.17$ \\
                       & Male (N=17)& $1.50 \pm 0.00$ & $\mathbf{0.55 \pm 0.16}$ & $2.30 \pm 0.26$ & $1.50 \pm 0.00$ \\ \hline
\multicolumn{6}{c}{\textbf{Marginal Groups (Gender Only)}} \\ \hline
\multirow{2}{*}{All}   & Fem (N=24) & $1.01 \pm 0.06$ & $\mathbf{0.64 \pm 0.05}$ & $0.93 \pm 0.05$ & $0.83 \pm 0.08$ \\
                       & Male (N=30)& $0.88 \pm 0.00$ & $\mathbf{0.40 \pm 0.10}$ & $1.05 \pm 0.15$ & $0.88 \pm 0.00$ \\ \hline
\multicolumn{6}{c}{\textbf{Marginal Groups (Age Only)}} \\ \hline
0-30                   & All (N=13) & $0.85 \pm 0.09$ & $\mathbf{0.50 \pm 0.00}$ & $0.58 \pm 0.09$ & $0.78 \pm 0.08$ \\
30-45                  & All (N=14) & $0.86 \pm 0.00$ & $\mathbf{0.63 \pm 0.17}$ & $1.17 \pm 0.13$ & $0.74 \pm 0.06$ \\
45+                    & All (N=27) & $1.20 \pm 0.00$ & $\mathbf{0.44 \pm 0.08}$ & $1.20 \pm 0.09$ & $1.08 \pm 0.10$ \\ \hline
\end{tabular}
}
\end{table}



\section{Conclusions}

In this study, we propose a methodological framework to evaluate transparent and robust speech-based depression detection pipelines on the extended DAIC-WOZ corpus using low-complexity models (RF, SVM, MLP) and human-understandable acoustic features. Crucially, we emphasize a selection–validation protocol—combining cost-sensitive evaluation (NEC) with permutation and consistency tests plus statistical and demographic analyses—to filter out spurious, artifact-driven cues and retain effects that replicate across runs and conditions.
Notably, Test B—which compares the distributions of systems utilizing the interviewer's voice against those using only the participant's voice—was successfully passed by the majority of the models. This indicates that, at least under the proposed framework, the system does not exploit shortcuts to improve its metrics using the interviewer's voice.
The utility of this framework is demonstrated by our MLP guided by LIME-based feature selection, which delivers the most reliable gains over baseline with SOTA metrics of 82\% accuracy on test and statistical significance achieve throughout the tests. The resulting explanations consistently point to plausible paralinguistic correlates of depression (e.g., breathiness/prosodic instability, longer response latencies, reduced articulatory dynamics reflected in temporal MFCC patterns), aligning with clinical notions such as psychomotor retardation and showing that the LIME feature extraction procedure is highly interpretable.

Finally, our fairness analysis shows that strong overall performance does not guarantee equitable utility: even top systems exhibit cohort-dependent disparities (e.g., better performance for younger/middle-aged males). These gaps—likely amplified by limited and imbalanced subgroup samples—reinforce the need for the proposed framework’s interpretability and rigorous validation as prerequisites for auditing, trust, and responsible deployment.
\section*{Acknowledgments}

This work has received funding from the European Union’s Horizon Europe research and innovation programme under the Marie Skłodowska-Curie grant agreement No 101206575, MCIN/AEI/10.13039/501100011033 under Grant PID2024-155948OB-C53.
The authors acknowledge the use of GitHub Copilot \cite{githubcopilot} to assist in the refinement and debugging of the Python scripts utilized in Section II. Additionally, Google's Gemini \cite{gemini2023} was utilized for language polishing and grammar correction of the manuscript. All AI-generated code fragments and text modifications were thoroughly reviewed, manually verified, and integrated into the final framework by the authors.

\bibliographystyle{IEEEtran}
\bibliography{mybib}

\begin{thebibliography}{00}
\bibitem{b1} G. Eason, B. Noble, and I. N. Sneddon, ``On certain integrals of Lipschitz-Hankel type involving products of Bessel functions,'' Phil. Trans. Roy. Soc. London, vol. A247, pp. 529--551, April 1955.
\end{thebibliography}

\end{document}